\documentclass[12pt]{article}

\usepackage{amssymb}
\usepackage{amsmath}
\usepackage{bm}
\usepackage{upgreek}
\usepackage{graphicx}

\begin{document}

\begin{center}
{\bf CASIMIR FRICTION FORCE BETWEEN POLARIZABLE MEDIA}

\vspace{1cm}
Johan S. H{\o}ye\footnote{johan.hoye@ntnu.no}

\bigskip
Department of Physics, Norwegian University of Science and Technology, N-7491 Trondheim, Norway

\bigskip
Iver Brevik\footnote{iver.h.brevik@ntnu.no}

\bigskip
Department of Energy and Process Engineering, Norwegian University of Science and Technology, N-7491 Trondheim, Norway

\bigskip
%\today
\end{center}

\begin{abstract}

This work is a continuation of our recent series of papers on Casimir friction, for a pair of particles of low relative particle velocity. Each particle is modeled as a simple harmonic oscillator. Our basic method, as before, is the use of quantum mechanical statistical mechanics, involving the Kubo formula, at finite temperature. In this work  we begin by analyzing the Casimir friction between two particles polarizable in all spatial directions, this being a generalization of our study in EPL {\bf 91}, 60003 (2010), which was restricted to a pair of particles with  longitudinal polarization only. For simplicity the particles are taken to interact via the electrostatic dipole-dipole interaction.  Thereafter, we consider the Casimir friction between one particle and a dielectric half-space, and also the friction between two dielectric half-spaces. Finally, we consider general polarizabilities (beyond the simple one-oscillator form),   and show how friction occurs at finite temperature when finite frequency regions  of  the imaginary parts of polarizabilities    overlap.

\end{abstract}


\section{Introduction}

Casimir friction - a subclass of the Casimir field of research - has emerged to be a topic of considerable current interest. It is basically a non-contact kind of friction that may be related to electromagnetic fluctuations. The effect is small under normal circumstances, and may moreover be difficult to deal with theoretically since it involves energy dissipation, necessitating in turn the use of complex valued permittivities in macroscopic electrodynamics. Once one leaves the state of thermal equilibrium one will have to face fundamental problems. Thus, in connection with macroscopic electrodynamics a very instructive way of approach is to make use of the spectral summation method, but this method is based upon real eigenfrequencies in the spectral problem and how to deal with such a system in the case of dissipation is at present unclear. Cf., for instance, the discussion on this point in Refs.\cite{bordag11} and \cite{nesterenko11}.

Most of the previous works in this area are based on the macroscopic dielectric model with permittivity properties. See, for instance, Refs.~\cite{levitov89,pendry97,pendry10,volokitin08,dedkov08,dedkov10,philbin09}. There exists, however, a different strategy in order to deal with Casimir friction, namely to consider the statistical mechanics for harmonic oscillators at finite temperature $T$ moving with constant velocity $v$ relative to each other. We argue that such a microscopic model, in spite of its simplicity compared with a full-fledged macroscopic model, has nevertheless the capacity of providing physical insight in the problem. And this is the kind of model that we shall consider in the following. We shall make an extension of our considerations in recent papers \cite{hoye10,hoye10a,hoye11}; cf. also the arXiv report \cite{hoye11b}. And this is again based upon a study of ours back in 1992 \cite{hoye92},
 dealing with the same kind of system. It contains a generalization to the time-dependent case of the statistical mechanical Kubo formalism spelled out for the time-independent case in Ref.~\cite{brevik88}.

Whereas our previous works were limited to the case of two microscopic oscillators, we shall here generalize the formalism so as to deal with one oscillator outside a slab (a collection of oscillators). Also, the generalization to two slabs in relative constant motion is easily achievable.

We emphasize that we are considering dilute media only.  The generalization to media of arbitrary density ought to be tractable with the actual method, but has to our knowledge not been treated yet. Our results are that the energy change $\Delta E$ because of Casimir friction is finite in general. This corresponds to a finite friction force. At zero temperature the formalism yields $\Delta E \rightarrow 0$, however; this being due to our assumption of a slowly varying coupling.  For rapidly varying couplings, there will be a finite friction force also at $T=0$ \cite{hoye10a,barton10}.

We mention finally that the microscopic approach has been analyzed by other investigators also, especially by Barton in recent papers \cite{barton10,barton10a,barton11}. The equivalence between Barton's results and our own results is not so easy to see by mere inspection since the methods are different, but were shown explicitly to be equivalent in one of our recent papers \cite{hoye11}.

\section{Friction force between two oscillators}

Consider the quantum mechanical system of two polarizable particles whose reference state is that of uncoupled motion corresponding to a Hamiltonian $H_0$. The equilibrium situation is then perturbed by a time independent term $-AF(t)$ where $A$ is a time independent operator and $F(t)$ is a classical time dependent  function. For a pair of polarizable particles perturbed with  dipole-dipole interaction we have
\begin{equation}
-AF(t)=\psi_{ij}s_{1i}s_{2j}, \label{1}
\end{equation}
where the summation convention for repeated indices $i$ and $j$ is implied. Here $s_{1i}$ and $s_{2j}$ are the components of the fluctuation dipole moments of the two particles $(i,j=1,2,3)$. With the electrostatic dipole-dipole interaction,
\begin{equation}
\psi_{ij}=-\left( \frac{3x_i x_j}{r^5}-\frac{\delta_{ij}}{r^3}\right), \label{2}
\end{equation}
where ${\bf r}={\bf r}(t)$ with components $x_i=x_i(t)$ is the separation between the particles. In an earlier work we also studied the situation with time-dependent interaction (with retardation effects) \cite{hoye93}, but we will avoid this added complexity here.

The situation with which we shall  mainly be concerned in the following is when the relative velocity $\bf v$ of the oscillators is constant. Then the interaction will vary as
\begin{equation}
-AF(t)=\left[ \psi_{ij}({\bf r}_0)+\left(\frac{\partial}{\partial x_l}\psi_{ij}({\bf r}_0)\right)v_l t+...\right]s_{1i}s_{2j}, \label{3}
\end{equation}
and the components of the force $\bf B$ between the oscillators are
\begin{equation}
B_l=-\frac{\partial}{\partial x_l}\left(\psi_{ij}s_{1i}s_{2j}\right). \label{4}
\end{equation}
The equilibrium situation with both particles at rest is represented by the first term in (\ref{2}). It gives rise to the (reversible) equilibrium force. Thus the friction is connected with the second term. To simplify, we shall here neglect the first term by which the two oscillators will be fully uncorrelated in their in their relative position ${\bf r}={\bf r}_0$. Thus we can write $-AF(t)\rightarrow -A_lF_l(t)$ where $A_l=B_l$ and $F_l(t)=v_lt.$ The friction force will be a small perturbation upon the equilibrium situation, and it leads to a response $\Delta \langle B_l(t)\rangle$ in the thermal average of $B_l$. According to Kubo \cite{kubo59,brevik88,hoye92}
\begin{equation}
\Delta \langle B_l(t)\rangle=\int_{-\infty}^\infty \phi_{BAlq}(t-t')F_q(t')dt', \label{5}
\end{equation}
where the response function is
\begin{equation}
\phi_{BAlq}(t)=\frac{1}{i\hbar}{\rm{Tr}}\, \rho[A_q, B_l(t)]. \label{6}
\end{equation}
Here $\rho$ is the density matrix and $B_l(t)$ is the Heisenberg operator $B_l(t)=e^{itH/\hbar}B_l\,e^{-itH/\hbar}$ where $\beta$ like $A$ is time independent. With Eqs. (\ref{3}) and (\ref{4}) and with $F_l (t)=v_lt$ expression (6) can be rewritten as
\begin{equation}
\phi_{BAlq}(t)=G_{lqijnm}\phi_{ijnm}(t), \label{7}
\end{equation}
where
\begin{equation}
G_{lqijnm}=\frac{\partial \psi_{ij}}{\partial x_l} \,\frac{\partial \psi_{nm}}{\partial x_q}, \label{8}
\end{equation}
\begin{equation}
\phi_{ijnm}(t)={\rm Tr} \{\rho C_{ijnm}(t)\}, \label{9}
\end{equation}
\begin{equation}
C_{ijnm}(t)=\frac{1}{i\hbar}\left[ s_{1i}s_{2j}, s_{1n}(t) s_{2m}(t)\right] \label{10}
\end{equation}
(the $i$ in the denominator is the imaginary unit).

Here as in Refs.~\cite{hoye92} and \cite{hoye10} the perturbing interaction (\ref{1}) will be considered weak. This will also hold for dilute dielectric media by which thermal averages of products, containing $s_{1i}$ and $s_{2j}$ as factors,  to leading order factorize. Further, with isotropy or scalar polarizability the components of each of the fluctuating dipole moments are uncorrelated.

From now on we find it convenient to utilize imaginary time, which was used in Sec.~4 of Ref.~\cite{hoye92}. With this we get
\begin{equation}
g_{ijnm}(\lambda)={\rm Tr}[\rho s_{1n}(t)s_{2m}(t)s_{1i}s_{2j}]=g_{1in}(\lambda)g_{2jm}(\lambda), \label{11}
\end{equation}
\begin{equation}
g_{apq}(\lambda)=\langle s_{aq}(t)s_{ap}\rangle =g_a(\lambda)\delta_{qp}, \quad (a=1,2), \label{12}
\end{equation}
where $g(\lambda)$ is the correlation function and
 angular brackets denote thermal averages $(\langle ..\rangle={\rm Tr}[\rho...])$. The $\lambda$ is imaginary time given by
\begin{equation}
\lambda=i\frac{t}{\hbar}, \label{13}
\end{equation}
so that for an operator $B$
\begin{equation}
B(t)=e^{\lambda H}B\,e^{-\lambda H}. \label{14}
\end{equation}
With this,
\[ \phi_{ijnm}(t)=\frac{1}{i\hbar}\left[ g_{ijnm}(\beta+\lambda)-g_{ijnm}(\lambda)\right] \]
\begin{equation}
=\frac{1}{i\hbar}\left[ g_1(\beta+\lambda)g_2(\beta+\lambda)-g_1(\lambda)g_2(\lambda)\right]\delta_{in}\delta_{jm}. \label{15}
\end{equation}
The response function $\phi_{ijnm}(t)$ corresponds to the retarded Green function, in the usual language of quantum field theory.

Earlier we unfortunately made a mistake by defining $\lambda$ in Eq.~(\ref{13}) with opposite sign \cite{hoye92}. However, this did not influence the results of previous applications since the operators $A$ and $B$ were equal apart from prefactors.

In Appendix B of Ref.~\cite{hoye92} it was found that
\begin{equation}
\tilde{\phi}(\omega)=\tilde{g}(K), \label{16}
\end{equation}
so that the Fourier transforms of the response function $\phi$ and the correlation function $g$ are equal. Explicitly,
\[ \tilde{\phi}(\omega)=\int_0^\infty \phi(t)e^{-i\omega t}dt \quad (\phi(t)=0~{\rm for} ~t<0), \]
\begin{equation}
\tilde{g}(K)=\int_0^\beta g(\lambda)e^{iK\lambda} d\lambda, \label{17}
\end{equation}
and
\begin{equation}
K=i\hbar \omega. \label{18}
\end{equation}
[ In Ref.~\cite{hoye92} $K=-i\hbar \omega$ was used due to the mistake mentioned.] Equality (\ref{16}) holds in the common region $|{\rm Im}(K)|<C$ with $C>0$ and ${\rm Im} (\omega)<0$, i.e., ${\rm Re} (K)>0$ where both functions are analytic.

It may be remarked that Eqs.~(\ref{12})- (\ref{18}) above correspond in a quantum field theoretical language to the statement that the spectral correlation function (frequency $\omega$) is equal to the imaginary part of the spectral retarded Green function multiplied with $\coth(\frac{1}{2} \beta\hbar \omega)$. See, for instance, Eq.~(76.6) in Ref.~\cite{landau81}. In the present case the function $\tilde g_+(\omega)$ which is the real time Fourier transform of $g_+(t)=\frac{1}{2\hbar}(g(it/\hbar)+g(-it/\hbar))$, will be the spectral correlation function while the $\tilde\phi(-\omega)$ (with Fourier transform (\ref{17})) will be the spectral retarded Green function (for $\omega>0$ as $\tilde g_+(\omega)$ is symmetric in $\omega$ while Im$[\tilde\phi(-\omega)]$ is antisymmetric).

With Eqs.~(\ref{11}) and (\ref{12}) we have
\[ g_{ijnm}(\lambda)=g(\lambda)\delta_{in}\delta_{jm}, \]
\begin{equation}
g(\lambda)=g_1(\lambda)g_2(\lambda). \label{19}
\end{equation}
Thus in $K$- space the ${\tilde g}(K)$ can be written as the convolution
\begin{equation}
{\tilde g}(K)=\frac{1}{\beta}\sum_{K_0}{\tilde g}_1(K_0){\tilde g}_2(K-K_0), \label{20}
\end{equation}
which is Eq.~(4.15) of Ref.~\cite{hoye92}.

An advantage of using imaginary time is that ${\tilde g}_a(K)$ can be identified with the frequency dependent polarizability $\alpha_{aK}$ of oscillator $a~(=1,2)$. For a simple harmonic oscillator with eigenfrequency $\omega_a$ one has \cite{brevik88}
\begin{equation}
{\tilde g}_a(K)=\alpha_{aK}=\frac{\alpha_a(\hbar \omega_a)^2}{K^2+(\hbar \omega_a)^2}, \label{21}
\end{equation}
where $\alpha_a$ is the zero-frequency polarizability. [ The ${\tilde g}_i(K)$ given by Eq.~(4.10) of Ref.~\cite{hoye92} will differ from the one of Eq.~(\ref{21}) by a factor $e^2$ where $e$ is the electron charge since here the ${\bf s}_a$ is identified with a dipole moment.]

With the above one finds for the Fourier transform of (\ref{9})
\begin{equation}
{\tilde \phi}_{ijnm}(\omega)={\tilde \phi}(\omega)\delta_{in}\delta_{jm}, \label{22}
\end{equation}
with ${\tilde\phi}(\omega)={\tilde g}(K).$ This follows by use of Eqs.~(\ref{15}) and (\ref{19}) from which ${\tilde g}_{ijnm}(K)={\tilde g}(K)\delta_{in}\delta_{jm}$, and when account is taken of Eq.~(\ref{16}). The ${\tilde g}(K)$ is moreover given by the convolution (\ref{20}) where for two simple harmonic oscillators the ${\tilde g}_a(K)~(a=1,2)$ are given by Eq.~(\ref{21}).

To obtain the perturbing force (\ref{5}) the expression (\ref{7}) should be evaluated. With (\ref{22}) one can write
\begin{equation}
\phi_{ijnm}(t)=\phi(t)\delta_{in}\delta_{jm}, \label{23}
\end{equation}
by which
\begin{equation}
\phi_{BAlq}(t)=G_{lq}\,\phi(t), \label{24}
\end{equation}
\begin{equation}
G_{lq}=G_{lqijij}=T_{lij} T_{qij}, \label{25}
\end{equation}
\begin{equation}
T_{lij}=\frac{\partial \psi_{ij}}{\partial x_l}. \label{26}
\end{equation}
With $\psi_{ij}$ given by Eq.~(\ref{2}) one finds
\begin{equation}
T_{lij}=\frac{15x_ix_jx_l}{r^7}-\frac{3(x_i\delta_{lj}+x_j\delta_{il}+x_l\delta_{ij})}{r^5}, \label{27}
\end{equation}
by which
\[ G_{lq}=T_{lij}T_{qij}=\left[ \left(\frac{15}{r^7}\right)^2r^4-\frac{270}{r^{12}}r^2+\left(\frac{9}{r^5}\right)^2 \right] x_lx_q \]
\begin{equation}
+\left(\frac{3}{r^5}\right)^2 2r^2 \delta_{lq}=\frac{18}{r^8}\delta_{lq}+\frac{36x_lx_q}{r^{10}}. \label{28}
\end{equation}
With the above expressions the result for the perturbing force (\ref{5}) will be precisely the same as obtained in Ref.~\cite{hoye10}, except from subscripts $l$ and $q$. One may simply insert Eq.~(\ref{24}) into Eq.~(\ref{5}) with $G_{lq}$ given by Eq.~(\ref{28}), $\phi(t)$ following from the relations (\ref{16})-(\ref{23}).

Thus, like in Eqs.~(8) and (9) of Ref.~\cite{hoye10} we here get the perturbing force
\begin{equation}
F_l=\Delta \langle B_l(t)\rangle = F_{rl}+F_{fl}, \label{29}
\end{equation}
where
\begin{equation}
F_{rl}=G_{lq}v_qt\int_0^\infty \phi(u)du \label{30}
\end{equation}
is part of the reversible force. The part of the force representing friction is
\begin{equation}
F_{fl}=-G_{lq}v_q\int_0^\infty \phi(u)udu. \label{31}
\end{equation}
The Fourier transformed version of this equation is, like Eq.~(11) of Ref.~\cite{hoye10}),
\begin{equation}
F_{fl}=-iG_{lq}v_q\frac{\partial {\tilde \phi}(\omega)}{\partial \omega}\Big|_{\omega=0}. \label{32}
\end{equation}
The expression for $\tilde{\phi}(\omega)$ is evaluated below Eq.~(\ref{38}) in the next section, by which the explicit expression (\ref{43}) is obtained for $F_{fl}$.

 Now the dissipated energy may be obtained; this requiring a perturbing interaction to last a finite amount of time to make it unique. Thus the force expression may be modified to
\begin{equation}
F_l(t) \rightarrow q_l(t), \label{33}
\end{equation}
where $q_l(t)$ is interpretable as a position. The dissipated energy $\Delta E_d$ is then given by expression (27) of Ref.~\cite{hoye10}. In the present case this will be $(A_l=B_l)$
\begin{equation}
\Delta E_d=\int_{-\infty}^\infty \dot{q}_l(t)\left[\int_{-\infty}^t \phi_{AAlq}(t-t')q_q(t')dt'\right] dt. \label{34}
\end{equation}

\section{Friction between particle and a half-space, and between two half-spaces}

Armed with the result (\ref{28}) for $G_{lq}$ it is now straightforward to extend the results in the previous section to the situation where a polarizable particle moves parallel to a resting dielectric half-space. We assume then that the particle density $\rho$ in the half-space is low so that the forces are additive. Let the half-space be located at $z\geq z_0$ such that its surface is parallel to the $xy$ plane at the vertical position $z=z_0$. The dielectric particle is located at the origin and moves with constant velocity $v_x=v_1$ along the $x$ axis. The resulting friction then follows by integrating $G_{lq}$ with $l=q=1$ over the dielectric half-space. With expression (\ref{28}) we find
\begin{equation}
G_h=\rho\int_{z\geq z_0}G_{11}dxdydz. \label{35}
\end{equation}
Symmetry with respect to the $x$ and $y$ coordinates means that the $x^2$ in Eq.~(\ref{28}) can be replaced with $\frac{1}{2}(x^2+y^2)$, and we can use cylindrical coordinates with $\rho^2=x^2+y^2$ and $dxdy=2\pi \rho d\rho$. Thus with $r^2=\rho^2+z^2$,
\begin{equation}
 \int G_{11} dxdy=36\pi \int_0^\infty \left( \frac{1}{r^8}+\frac{\rho^2}{r^{10}}\right) \rho d\rho =\frac{15\pi}{2z^6}, \label{36}
 \end{equation}
 by which
 \begin{equation}
 G_h=\rho\int_{z_0}^\infty \frac{15\pi}{2z^6}dz=\frac{3\pi \rho}{2z_0^5}. \label{37}
 \end{equation}
 Finally, for two dielectric half-spaces moving parallel relative to each other one can obtain the friction force per unit area from
 \begin{equation}
 G=\rho_2\int_{d}^\infty G_{h}dz_0=\frac{3\pi}{8d^4}\rho_1\rho_2, \label{38}
 \end{equation}
 where $d$ is the gap width and  $\rho_1$ and $\rho_2$ are the (low) particle densities in the two half spaces.

 The function $\tilde{\phi}(\omega)=\tilde{g}(K)$ is needed to obtain the friction. Performing the summation in Eq.~(\ref{20}), inserting the expression (\ref{21}) we obtain, in agreement with Eq.~(4.16) in \cite{hoye92},
 \begin{equation}
 \tilde{g}(K)=Hf(K), \label{39}
 \end{equation}
 with
 \[ H=\frac{E_1E_2\alpha_1\alpha_2}{4\sinh(\frac{1}{2}\beta E_1)\sinh(\frac{1}{2}\beta E_2)}, \]
 \begin{equation}
 f(K)=\frac{\Sigma_1\sinh(\frac{1}{2}\beta\Sigma_1)}{K^2+\Sigma_1^2}+\frac{\Sigma_2\sinh(\frac{1}{2}\beta\Sigma_2)}{K^2+\Sigma_2^2}. \label{40}
 \end{equation}
 Here $\Sigma_1$ and $\Sigma_2$ are defined as
 \[ \Sigma_1=E_1+E_2, \quad \Sigma_2=E_1-E_2, \quad E_i=\hbar \omega_i~(i=1,2). \]
 When the velocity is small and constant (or very slowly varying), only the limit $K\rightarrow 0$ is needed. One further sees that the contribution requires that $\Sigma_2\rightarrow 0$. Because of this, the $f(K)$ becomes a $\delta$-function (plus a constant)
 \begin{equation}
 f(K)=-\frac{\pi}{2}\beta K\delta(\Sigma_2), \label{41}
 \end{equation}
 like Eq.~(4.18) of Ref.~\cite{hoye92}. [Here Re$(K)>0$ in view of the correction mentioned below Eq.~(\ref{15}).] To obtain the friction force like (\ref{32}) for a pair of particles and thus  the more general situation, the derivative of $f(K)$ with respect to $\omega$ is needed. With $\delta(\Sigma_2)=\delta(\hbar(\omega_1-\omega_2))=\delta(\omega_1-\omega_2)/\hbar$ and $K=i\hbar \omega$ one finds
 \begin{equation}
 i\frac{\partial}{\partial \omega}f(K)=-\hbar \frac{\partial f(K)}{\partial K}=\frac{\pi}{2}\beta \delta(\omega_1-\omega_2). \label{42}
 \end{equation}
 For a pair of polarizable particles the friction force (\ref{32}) thus becomes
 \begin{equation}
 F_{fl}=-G_{lq}v_qH \frac{\pi \beta}{2}\delta(\omega_1-\omega_2), \label{43}
 \end{equation}
 with $H$ given by Eq.~(\ref{40}) and $G_{lq}$ by Eq.~(\ref{28}).

 For the more simple model studied in Ref.~\cite{hoye10} the $G_{lq}v_q$ is replaced by ${\bf G}= ({\bf \nabla}\psi)({\bf v}\cdot {\bf \nabla}\psi)$. The result (19)  of that reference is recovered if the $(\hbar \omega_a)^2\alpha_a~(a=1,2)$ is replaced with $\hbar^2/m_a$ in the expression (\ref{40}) for $H$. This replacement follows from a corresponding change in $\tilde{g}_a(K)$ as given by Eq.~(\ref{21}) to the one given by Eq.~(4.10) in Ref.~\cite{hoye92}.

 It is now straightforward to obtain the friction $F_h$ between a polarizable particle and a half-space. One can simply replace the $G_{lq}v_q$ in Eq.~(\ref{43}) with $G_h v$, where $v$ is the velocity of the particle parallel to the plane and $G_h$ is given by expression (\ref{37}). Thus
 \begin{equation}
 F_h=-\frac{3\pi \rho}{2z_0^5}H\frac{\pi \beta}{2}\delta(\omega_1-\omega_2). \label{44}
 \end{equation}
 Likewise, with two half-spaces moving relative to each other the $G_h$ is replaced with $G$ given by Eq.~(\ref{38}) to obtain the friction force per unit area
 \begin{equation}
 F=-\frac{3\pi}{8d^4}\rho_1\rho_2 H\frac{\pi \beta}{2}\delta(\omega_1-\omega_2). \label{45}
 \end{equation}

 \section{General polarizability}

 For simple harmonic oscillators the polarizability is given by Eq.~(\ref{21}). However, it can be a more general function of $K$ that may be regarded as resulting from a sum of harmonic oscillators. Thus we may write
 \begin{equation}
 h(K^2)=\tilde{g}_a(K)=\alpha_{aK}, \label{46}
 \end{equation}
 where it can be shown \cite{hoye82} that the function $h(K^2)$ satisfies the relation
 \begin{equation}
 h(K^2)=\int \frac{\alpha_a(m^2)m^2}{K^2+m^2}d(m^2), \label{47}
 \end{equation}
 with
 \begin{equation}
 \alpha_a(m^2)m^2=-\frac{1}{\pi}{\rm Im}[h(-m^2+i\gamma)], \quad (m=\hbar \omega=-iK,~\gamma \rightarrow 0^+). \label{48}
 \end{equation}
 This is analogous to the Kramers-Kronig relations for the dielectric permittivity. An implication of the more general expression for $\tilde{g}_a(K)$ is that the $\alpha_a$ $(a=1,2)$ in Eq.~(\ref{40}) is replaced with
  $\int \alpha_a(m_a^2)d(m_a^2)$. Integration of the $\delta$-function in Eqs.~(\ref{42})-(\ref{45}) further gives
 \begin{equation}
 \int\delta(\omega_1-\omega_2)d(m_1^2)d(m_2^2)=4(\hbar \omega_1)^2 d\omega_1. \label{49}
 \end{equation}
 So altogether the $H(\pi \beta/2) \delta(\omega_1-\omega_2)$ in these equations will be replaced by $(\omega_1,\omega_2\rightarrow \omega, m=\hbar \omega)$
 \begin{equation}
 H_0=\frac{\pi \beta}{2}\int \frac{m^4\alpha_1(m^2)\alpha_2(m^2)}{\sinh^2(\frac{1}{2}\beta m)} d\omega. \label{50}
 \end{equation}
 With this the friction forces, Eqs.~(\ref{43})-(\ref{45}), for a pair of particles, one particle and a half-space, and two half-spaces, become respectively
 \[ F_{fl}=-G_{lq}v_qH_0, \]
 \[ F_h=-\frac{3\pi}{2}\frac{\rho}{z_0^5}H_0, \]
 \begin{equation}
 F=-\frac{3\pi}{8d^4}\rho_1\rho_2 H_0. \label{51}
 \end{equation}
 Thus, at a finite temperature a finite friction force is obtained when finite values of $\alpha_1(m^2)$ and $\alpha_2(m^2)$, as defined in Eqs.~(\ref{46})-(\ref{48}),  overlap.

\end{document}